\documentclass[seceq]{ptptex}

\usepackage{graphicx}

\def\be{\begin{equation}}
\def\ee{\end{equation}}
\def\bea{\begin{eqnarray}}
\def\eea{\end{eqnarray}}

\title{
On the width of collective excitations in chiral soliton models
}

\author{
Herbert \textsc{Weigel}\footnote{ e-mail address:
Weigel@physik.uni-siegen.de}  }

\inst{
Fachbereich Physik, Siegen University,
D--57068 Siegen, Germany}

\abst{
In chiral soliton models for baryons the computation of hadronic decay 
widths of baryon resonances is a long standing problem. For the three flavor 
Skyrme model I present a solution to this problem that satisfies large--$N_C$ 
consistency conditions. As an application I focus on the hadronic decay of 
the $\Theta$ and $\Theta^*$ pentaquarks.}

\begin{document}

\maketitle

\section{Statement of the problem}

Hadronic decays of baryon resonances are commonly described 
by a Yukawa interaction of the generic structure
\begin{equation}
{\mathcal L}_{\rm int} \sim g\,\bar{\psi}_{B^\prime}\,\phi\,\psi_B\,,
\label{eq:yukawa}
\end{equation}
where $B^\prime$ is the resonance that decays into baryon $B$ and meson 
$\phi$ and $g$ is a coupling constant. It is crucial that this 
interaction Lagrangian is \emph{linear} in the meson field.
If $\phi$ is a pseudoscalar meson this interaction yields the 
decay width $\Gamma(B^\prime\to B\phi)\propto g^2 |\vec{p}_\phi|^3$,
with $\vec{p}_\phi$ being the momentum of the outgoing meson.

The situation is significantly different in soliton models that are based
on action functionals of only meson degrees of freedom, 
$\Gamma=\Gamma[\Phi]$. These action functionals contain
classical (static) soliton solutions, $\Phi_{\rm sol}$, that are 
identified as baryons. The interaction of these baryons with mesons
is described by the (small) meson fluctuations about the 
soliton: $\Phi=\Phi_{\rm sol}+\phi$. By pure definition we have
\begin{equation}
\frac{\delta \Gamma[\Phi]}{\delta \Phi}\Big|_{\Phi=\Phi_{\rm sol}}=0\,.
\label{eq:soliton}
\end{equation}
Thus there is no term linear in $\phi$ to be associated with the Yukawa 
interaction, eq.~(\ref{eq:yukawa}). This puzzle has become famous as the 
Yukawa problem in soliton models. However, this does not mean that soliton 
models cannot describe resonance widths. On the contrary, these widths can be 
extracted from meson baryon scattering amplitudes, just as in experiment. In 
soliton models two--meson processes acquire contributions from the second 
order term 
\begin{equation}
\Gamma^{(2)}=\frac{1}{2}\,\phi \cdot\,
\frac{\delta^2 \Gamma[\Phi]}{\delta^2 \Phi}\Big|_{\Phi=\Phi_{\rm sol}}\,
\hspace{-0.3cm}\cdot\phi\quad .
\label{eq:secorder}
\end{equation}
This expansion simultaneously represents an expansion in $N_C$, the number 
of color degrees of freedom: $\Gamma=\mathcal{O}(N_C)$ and 
$\Gamma^{(2)}=\mathcal{O}(N_C^0)$. Terms $\mathcal{O}(\phi^3)$ vanish in 
the limit $N_C\to\infty$. This implies that $\Gamma^{(2)}$ contains all 
large--$N_C$ information about hadronic decays of resonances. We may reverse 
this statement to argue about any computation of hadronic decay widths in 
soliton models: For $N_C\to\infty$ it \emph{must} identically match the result 
obtained from $\Gamma^{(2)}$.  Unfortunately, the most prominent baryon resonance, 
the $\Delta$ isobar, becomes degenerate with the nucleon as $N_C\to\infty$. It is 
stable in that limit and its decay is not subject to the above described 
litmus--test. The situation is more interesting in soliton models for flavor 
$SU(3)$. In the so--called rigid rotator approach (RRA), that generates baryon 
states as (flavor) rotational excitations of the soliton, exotic resonances emerge 
that dwell in the anti--decuplet representation of flavor $SU(3)$\cite{Theta}. 
The most discussed (and disputed) such state is the $\Theta^+$ pentaquark with zero 
isospin and strangeness $S=+1$. In the limit $N_C\to\infty$ the (would--be) 
anti--decuplet states maintain a non--zero mass difference with respect to the 
nucleon. Therefore the properties of $\Theta^+$ predicted from any model treatment 
must also be seen in the $S$--matrix for koan--nucleon scattering as computed 
from~$\Gamma^{(2)}$. This (seemingly alternative) quantization of strangeness 
degrees of freedom is called the bound state approach (BSA) because in the $S=-1$ 
sector the resulting equations of motion for $\phi$  yield a bound state. Its 
occupation serves to describe the ordinary hyperons, $\Lambda$, $\Sigma$, $\Sigma^*$, 
etc.\@. The above discussed litmus--test requires that the BSA and RRA give identical 
results for the $\Theta^+$ properties as $N_C\to\infty$. This did not seem to be true 
and it was argued that the prediction of pentaquarks would be a mere artifact of the 
RRA~\cite{It04}. Here we will show that this is a premature conclusion and that 
pentaquark states do indeed emerge in both approaches. Furthermore the comparison 
between the BSA and RRA provides an unambiguous computation of pentaquark widths: 
It differs substantially from previous approaches~\cite{Di97} that adopted 
transition operators for $\Theta^+\to KN$ from the axial current.

This presentation is based on ref.~\cite{Wa05} which the interested 
reader may want to consult for further details.

\section{The model}

For simplicity we consider the Skyrme model~\cite{Sk61} as a 
particular example for chiral soliton models. However, we stress that our 
qualitative results do indeed generalize to \emph{all} chiral soliton models 
because these results solely reflect the treatment of the model 
degrees of freedom. 

Chiral soliton models are functionals of the chiral field, $U$, the 
non--linear realization of the pseudoscalar mesons\footnote{A remark 
on notation: In what follows we adopt the convention that repeated indices 
are summed over in the range $a,b,c,\ldots = 1,\ldots,8$, 
$\alpha,\beta,\gamma,\ldots = 4,\ldots,7$ and
$i,j,k,\ldots = 1,2,3$.}, $\phi_a$
\be
U(\vec{x\,},t)={\rm exp}\left[\frac{i}{f_\pi}
\phi_a(\vec{x\,},t)\lambda_a\right]\,,
\label{chiralfield}
\ee
with $\lambda_a$ being the Gell--Mann matrices of $SU(3)$. For a convenient 
presentation of the model we split the action into three pieces
\be
\Gamma = \Gamma_{SK} + \Gamma_{WZ} + \Gamma_{SB} \,.
\label{lag}
\ee
The first term represents the Skyrme model action
\be
\Gamma_{SK} =
\int d^4 x\, {\rm tr}\,\left\{ \frac{f^2_\pi}{4} 
\left[ \partial_\mu U \partial^\mu U^\dagger \right]
+ \frac{1}{32\epsilon^2} \left[ [U^\dagger \partial_\mu U , 
U^\dagger \partial_\nu U]^2\right] \right\} \, .
\label{Skmodel}
\ee
Here $f_\pi=93{\rm MeV}$ is the pion decay constant and $\epsilon$ is 
the dimensionless Skyrme parameter. In principle this is a free 
model parameter. The two--flavor version of the Skyrme model suggests to 
put $\epsilon=4.25$ from reproducing the $\Delta$--nucleon mass 
difference\footnote{To ensure that the (perturbative) $n$--point functions 
scale as $N_C^{1-n/2}$~\cite{tH74} we substitute 
$f_\pi=93{\rm MeV}\sqrt{N_C/3}$ and $\epsilon=4.25\sqrt{3/N_C}$ in the study 
of the $N_C$ dependence.}. The QCD anomaly is incorporated via the 
Wess--Zumino action~\cite{Wi83} 
\be
\Gamma_{WZ} = - \frac{i N_C}{240 \pi^2}
\int d^5x \ \epsilon^{\mu\nu\rho\sigma\tau}
{\rm tr}\,\left[\alpha_\mu\alpha_\nu\alpha_\rho\alpha_\sigma\alpha_\tau\right] \,,
\label{WZ}
\ee
with $\alpha_\mu = U^\dagger \partial_\mu U$. Note that 
$\Gamma_{WZ}$ vanishes in the two--flavor version of the model.

The flavor symmetry breaking terms are contained in $\Gamma_{\rm SB}$ 
\be
\Gamma_{SB} = \frac{f_\pi^2}{4}\, \int d^4x\,
{\rm tr}\, \left[{\cal M}\left(U + U^\dagger - 2 \right)\right]
\,,\quad
\mathcal{M}=\mbox{diag}\left(m_\pi^2,m_\pi^2 ,2m_K^2-m_\pi^2\right)\,.
\label{SB}
\ee
We do not include terms that distinguish between pion and kaon decay 
constants even though they differ by about 20\% empirically. This omission 
is a matter of convenience and leads to an underestimation 
of symmetry breaking effects~\cite{We90} which approximately can be 
accounted for by rescaling the kaon mass $m_K\to m_Kf_K/f_\pi$. 

The action, eq.~(\ref{lag}) allows for a topologically non--trivial
classical solution, the famous hedgehog soliton
\be
\Phi_{\rm sol}\,\sim\,U_0(\vec{x\,})
={\rm exp}\left[i\vec{\lambda\,}\cdot\hat{x} F(r)\right]\,,
\quad r=|\vec{x\,}|
\label{hedgehog}
\ee
embedded in the isospin subspace of flavor $SU(3)$. The chiral angle, $F(r)$ 
solves the classical equation of motion subject to the boundary condition
$F(0)-F(\infty)=\pi$ ensuring unit winding (baryon) number. The soliton 
can be constructed as a function of the dimensionless variable 
$\epsilon f_\pi r$ and is thus not subject to $N_C$ scaling.

In the RRA baryon states are generated by canonically quantizing collective
coordinates $A\in SU(3)$ that describe the (spin) flavor orientation
of the soliton, $A(t)U_0(\vec{x})A^\dagger(t)$. The resultant eigenstates
may be classified according to $SU(3)$ multiplets; see ref.~\cite{We96}
for a review.

\section{Large $N_C$ $P$--wave channel phase shifts with strangeness}

\begin{figure}[b]
\centering
\includegraphics[width=3.5cm,height=6cm,angle=270]{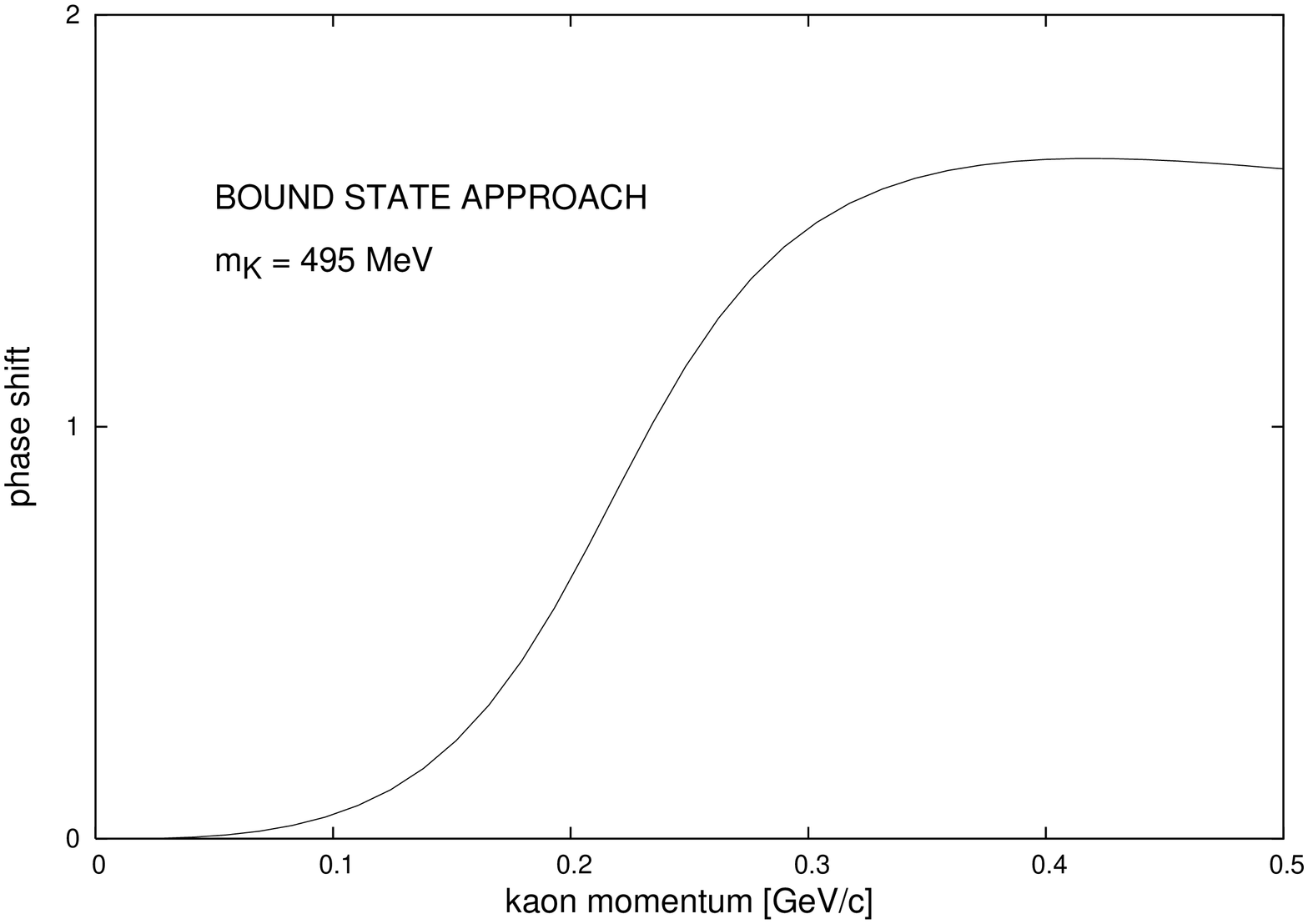}
\hspace{1cm}
\includegraphics[width=3.5cm,height=6cm,angle=270]{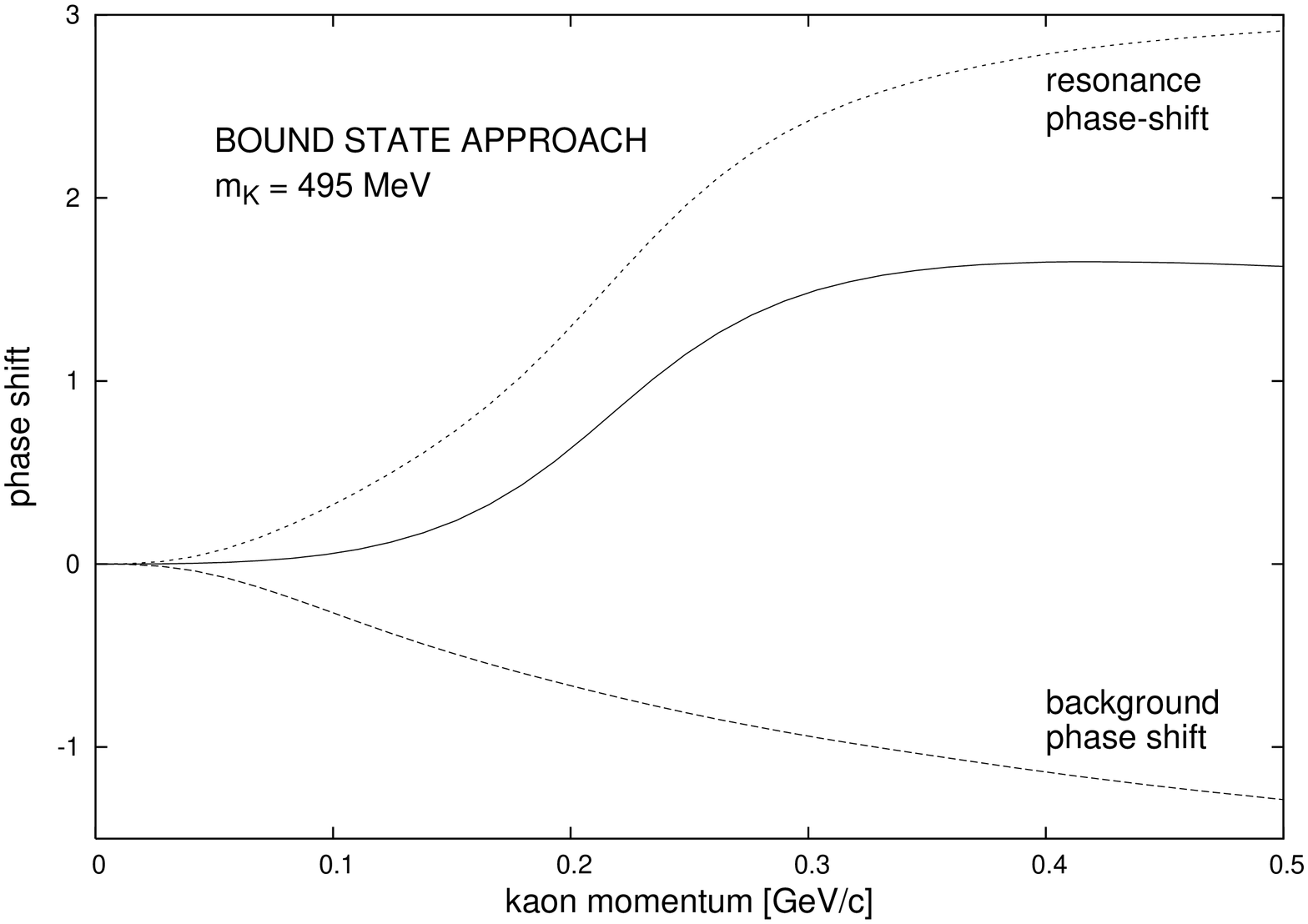}
\caption{\label{fig1}Large $N_C$ $P$--wave phase shifts with strangeness
$S=+1$ as function of the kaon momentum. 
Left panel: unconstrained; right panel: constrained to be
orthogonal to the collective rotation.}
\end{figure}

As motivated after eq.~(\ref{eq:secorder}) we introduce fluctuations 
$\phi\,\sim\,\eta_\alpha(\vec{x\,},t)$
\be
U(\vec{x\,},t)=  \sqrt{U_0(\vec{x\,})}\,
{\rm exp}\left[\frac{i}{f_\pi}\lambda_\alpha\eta_\alpha(\vec{x\,},t)\right]
\sqrt{U_0(\vec{x\,})}\,,
\label{ckfluct}
\ee
for the kaon fields~\cite{Ca85}.
Expanding the action in powers of these fluctuations is 
an expansion in $\eta_\alpha/f_\pi$ and thus a systematic series 
in $1/\sqrt{N_C}$. The term quadratic in $\eta_\alpha $ describes meson 
scattering off a potential generated by the classical soliton, 
eq.~(\ref{hedgehog}). The $P$--wave mode is characterized by a single 
radial function
\be
\begin{pmatrix}
\eta_4+i\eta_5\cr
\eta_6+i\eta_7
\end{pmatrix}_{P}(\vec{x\,},t)\,
=\int_{-\infty}^\infty\, d\omega\, {\rm e}^{i\omega t}\,
\eta_\omega(r)\,\hat{x}\cdot\vec{\tau}\,\chi(\omega)\,.
\label{pwave}
\ee
In future we will omit the subscript that indicates the Fourier 
frequency. Upon quantization the components of the two--component 
iso--spinor $\chi(\omega)$ are elevated to creation-- and annihilation 
operators. It is straightforward to deduce the Schr\"odinger type 
equation 
\be
h^2\,\eta(r)+\omega\left[2\lambda(r)-\omega M_K(r)\right]\eta(r)=0
\quad {\rm with} \quad 
h^2=-\frac{d^2}{dr^2}-\frac{2}{r}\frac{d}{dr}+V_{\rm eff}(r)\,.
\label{scndorder}
\ee
The radial functions arise from the chiral angle $F(r)$ and may
be readily taken from the literature~\cite{Ca85}. 
The equation of motion~(\ref{scndorder}) is not invariant under particle
conjugation $\omega\leftrightarrow-\omega$, and thus different for
kaons ($\omega>0$) and anti--kaons ($\omega<0$). This difference 
stems from the Wess--Zumino term.  Equation~(\ref{scndorder}) has a 
bound state solution at $\omega=-\omega_\Lambda$, that gives the mass 
difference between the 
$\Lambda$--hyperon and the nucleon in the large--$N_C$ limit. As this 
energy eigenvalue is negative it corresponds to a kaon, {\it i.e.\@} it 
carries strangeness $S=-1$. In the symmetric case ($m_K=m_\pi$) 
the bound state is simply the zero
mode of $SU(3)$ flavor symmetry. The WZ--term moves the potential 
bound state with $S=+1$ to the positive continuum and we expect 
a resonance structure in that channel. The corresponding phase shift
is shown in the left panel of figure~\ref{fig1}.
No clear resonance structure is visible; the phase shifts
hardly reach $\pi/2$. The absence of such a
resonance has previously lead to the premature criticism that
there would not exist a bound pentaquark in the
large--$N_C$ limit~\cite{It04}.

\section{Constraint fluctuations}

To study the coupling between the fluctuations and the collective
excitations we generalize eq.~(\ref{ckfluct}) to
\be
U(\vec{x\,},t)= A(t)\sqrt{U_0(\vec{x\,})}\,
{\rm exp}\left[\frac{i}{f_\pi}\lambda_\alpha
\widetilde{\eta}_\alpha(\vec{x\,},t)\right]
\sqrt{U_0(\vec{x\,})}A^\dagger(t)\,.
\label{ckfluctconst}
\ee
These fluctuations dwell in the intrinsic system as they rotate along with the 
soliton. The kaon $P$--wave is subject to the modified integro--differential 
equation
\bea
&&h^2\widetilde{\eta}(r)+
\omega\left[2\lambda(r)-\omega M_K(r)\right]\widetilde{\eta}(r)=
-z(r)\left[\int_0^\infty r^{\prime2}dr^\prime z(r^\prime)
2\lambda(r^\prime)\widetilde{\eta}(r^\prime)\right]\cr
&&\hspace{0.8cm}\times
\left[2\lambda(r)-\left(\omega+\omega_0\right)M_K(r)
-\omega_0\left(\frac{X_\Theta^2}{\omega_\Theta-\omega}
+\frac{X_\Lambda^2}{\omega}\right)
\left(2\lambda(r)-\omega_0M_K(r)\right)\right]\,,\hspace{1cm}
\label{Yukawa1}
\eea
for the flavor symmetric case\footnote{The more complicated case 
$m_K\ne m_\pi$ is at length discussed in ref.~\cite{Wa05}.}. The radial 
function $\widetilde{\eta}(r)$ is defined according to eq.~(\ref{pwave}) and 
$z(r)=\sqrt{4\pi}\frac{f_\pi}{\sqrt{\Theta_K}}\, {\rm sin}\frac{F(r)}{2}$ is
the collective mode wave--function normalized with respect to the moment of 
inertia for flavor rotations into strangeness direction, 
$\Theta_K=f_\pi^2 \int d^3r M_K(r)\,{\rm sin}^2\frac{F(r)}{2}=\mathcal{O}(N_C)$.
The non--local terms without $X_{\Lambda,\Theta}$ reflect the constraint 
$\int dr r^2 z(r)M_K(r)\widetilde{\eta}(r)=0$ which avoids double counting 
of rotational modes in strangeness direction. The interesting coupling
is contained in the interaction Hamiltonian
\be
H_{\rm int}=\frac{2}{\sqrt{4\pi\Theta_K}}\,
d_{i\alpha\beta}\, D_{\gamma\alpha}R_\beta\,
\int d^3r\, z(r)\left[2\lambda(r)-\omega_0 M_K(r)\right]
\hat{x}_i \widetilde{\xi}_\gamma(\vec{x\,},t)\,,
\label{hint}
\ee
where $\widetilde{\xi}_a=D_{ab}\widetilde{\eta}_b$ are the fluctuations in the 
laboratory frame, that we actually detect in $KN$ scattering. The collective 
coordinates are parameterized via the adjoint representation 
$D_{ab}(A)=\frac{1}{2}\,{\rm tr}\,\left[\lambda_a A \lambda_b A^\dagger\right]$
and the $SU(3)$ generators $R_a$. Integrating out the collective degrees of 
freedom by means of standard perturbation theory induces the separable potential
\be
\frac{\left|\langle\Theta|H_{\rm int}|(KN)_{I=0}\rangle\right|^2}
{\omega_\Theta-\omega}
+\frac{\left|\langle\Lambda|H_{\rm int}|(KN)_{I=0}\rangle\right|^2}
{\omega_\Lambda+\omega}\,.
\label{potential}
\ee
These matrix elements concern the $T$--matrix elements in the
laboratory frame. Since the laboratory and intrinsic $T$--matrix elements 
are identical for the $\Theta^+$ channel~\cite{Ha84}, we may 
add the exchange potential, eq.~(\ref{potential}) in the intrinsic frame.
We define matrix elements of collective coordinate operators
\be
\langle \Theta^+| d_{3\alpha\beta} D_{+\alpha}R_\beta|n\rangle 
=:X_\Theta\sqrt{\frac{N_C}{32}} \quad {\rm and}\quad
\langle \Lambda| d_{3\alpha\beta}
D_{-\alpha}R_\beta|p\rangle=:X_\Lambda\sqrt{\frac{N_C}{32}}\,,
\label{ddr}
\ee
to end up with eq.~(\ref{Yukawa1}). The first factor in the coefficient
$\omega_0=2\left(\frac{2}{\sqrt{\Theta_K}}\sqrt{\frac{N_C}{32}}\right)^2=
\frac{N_C}{4\Theta_K}$ arises in the equation of motion because the potential, 
eq.~(\ref{potential}) is quadratic in the fluctuations. The remaining 
(squared) factors stem from the definitions of $X_{\Theta,\Lambda}$ and the 
constant of proportionality in $H_{\rm int}$. The $X_{\Theta,\Lambda}$
must be computed with the methods provided in ref.~\cite{Ya88} but generalized 
to arbitrary (odd) $N_C$~\cite{Wa05}. For $N_C\to\infty$ we have
$X_\Theta\to1$ and $X_\Lambda\to0$. From the orthogonality conditions of the 
equation of motion~(\ref{scndorder}) we straightforwardly verify that
\be
\widetilde{\eta}(r)=\eta(r)-az(r)
\qquad {\rm with} \qquad
a=\int_0^\infty dr r^2\, z(r) M_K(r)\eta(r)\,.
\label{solution}
\ee
solves eq.~(\ref{Yukawa1}) for large $N_C$. This is essential
because, as $z(r)$ is localized in space, $\eta$ and $\widetilde{\eta}$
have identical phase shifts! Hence the litmus--test discussed in the introduction
is indeed satisfied. The physics behind $\widetilde{\eta}$ is best understood
when introducing reduced wave--functions $\overline{\eta}(r)$ that solve 
eq.~(\ref{Yukawa1}) modified with $X_\Theta\equiv X_\Lambda\equiv0$. 
{\it i.\@ e.\@} the collective excitations are decoupled. These
wave--functions are still orthogonal to the collective modes and actually
lead to the background phase shift shown in figure~\ref{fig1}. Having obtained
the $\overline{\eta}(r)$  we may again switch on the exchange contributions, 
eq.~(\ref{potential}). The additional separable potential is treated
by standard R--matrix techniques and augments the phase shift by
\be
{\rm tan}\left(\delta_{\rm R}(k)\right)=
\frac{\Gamma(\omega_k)/2}{\omega_\Theta-\omega_k+\Delta(\omega_k)}\,.
\label{resformula}
\ee
Here $\omega_\Theta=\frac{N_C+3}{4\Theta_K}$ is the RRA result for the 
excitation energy of $\Theta$.  This phase shift exhibits the canonical 
resonance structure with the width and the pole shift
\bea
\Gamma(\omega_k)&=&2k\omega_0 X_\Theta^2
\left|\int_0^\infty r^2dr\, z(r)2\lambda(r)
\overline{\eta}_{\omega_k}(r)\right|^2\,,
\label{width1} \\
\Delta(\omega_k)&=&\frac{1}{2\pi\omega_k}\,{\cal P}\,
\int_0^\infty q dq\,\left[
\frac{\Gamma(\omega_q)}{\omega_k-\omega_q}
+\frac{\Gamma(-\omega_q)}{\omega_k+\omega_q}\right]\,,
\label{poleshift}
\eea
respectively. We have numerically verified that in the large--$N_C$ limit 
with $X_\Theta^2=1$, the phase shift from eq.~(\ref{resformula}) is 
identical to what is labeled resonance phase shift in figure~\ref{fig1}, 
that we calculated as the difference between the total ($\eta$) and 
background ($\overline{\eta}$) phase shifts. For finite $N_C$ we have 
$X_\Theta\ne1$ and $X_\Lambda\ne0$ so the R--matrix formalism
becomes two--dimensional~($\Lambda$ and $\Theta^+$ exchange). Contrary to 
earlier criticisms~\cite{It04} the large $N_C$ pentaquark phase shift 
indeed resonates!

\section{Results}

\begin{figure}[t]
\centerline{
\includegraphics[width=3.5cm,height=10cm,angle=270]{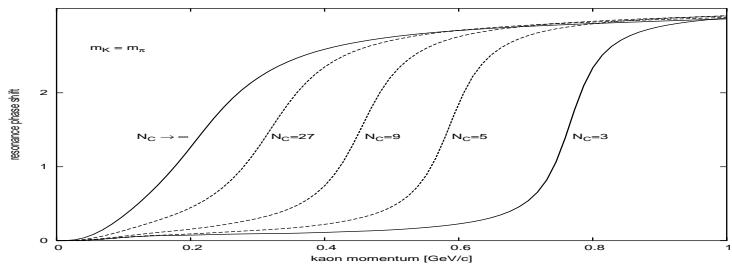}}
\caption{\label{fig2}The resonance phase shift as a function of $N_C$
for $m_K=m_\pi$.}
\end{figure}

In figure~\ref{fig2} we show the resonance phase shift computed from
eq.~(\ref{resformula}) for various values of $N_C$. First we observe that the 
resonance position quickly moves towards larger energies as $N_C$ decreases. This 
is mainly due to the strong $N_C$ dependence of $\omega_\Theta$: For $N_C=3$ it
is twice as large as in the limit $N_C\to\infty$. The pole shift $\Delta$
is actually quite small (some ten MeV) so that $\omega_\Theta$ is indeed
a reliable estimate of the resonance energy. Second, the resonance
becomes shaper as $N_C$ decreases. To major parts this is caused by
the reduction of $X_\Theta$.

We now turn to more quantitative results for which we also include 
flavor symmetry breaking effects. Then the resonance position changes to
\be
\omega_\Theta=\frac{1}{2}\left[\sqrt{\omega_0^2+\frac{3\Gamma}{2\Theta_K}}
+\omega_0\right]+\mathcal{O}\left(\frac{1}{N_C}\right)\,.
\label{omt}
\ee
where $\Gamma=\mathcal{O}(N_C)$ is a functional of the soliton that is 
proportional to the meson mass difference, $m_K^2-m_\pi^2$. The 
$\mathcal{O}\left(1/N_C\right)$ piece is sizable for $N_C=3$ and we 
compute it in the scenario of ref.\cite{Ya88}. We then find 
$\omega_\Theta\approx700{\rm MeV}$; taking model dependencies into
account we expect the pentaquark to be about $600\ldots900{\rm MeV}$ 
heavier than the nucleon.

For the width calculations there are two principle differences. 
First, the interaction Hamiltonian acquires an additional term 
\be
H_{\rm int}^{\rm sb}=
\left(m_K^2-m_\pi^2\right)d_{i\alpha\beta}D_{\gamma\alpha}D_{8\beta}
\int d^3r\,z(r)\gamma(r)\widetilde{\xi}_{\gamma}(\vec{x\,},t)\hat{x}_i\,,
\label{hintsb}
\ee
The radial function $\gamma(r)$ is again given in terms of the chiral 
angle~\cite{Wa05}. Second, the $X_\Lambda$ does not vanish as $N_C\to\infty$
and the $R$--matrix formalism is always two dimensional. Nevertheless, the 
large--$N_C$ solution is always of the form~(\ref{solution}) and the BSA
phase shift is recovered. The width function turns to
\be
\Gamma(\omega_k)=2k\omega_0 \left|\int_0^\infty \hspace{-1mm} r^2dr\, z(r)
\left[2X_\Theta\lambda(r) +\frac{Y_\Theta}{\omega_0}\left(m_K^2-m_\pi^2\right)
\right]\overline{\eta}_{\omega_k}(r)\right|^2,
\label{widthsb}
\ee
where $X_\Theta$ and $Y_\Theta=\sqrt{8N_C/3} 
\langle\Theta^+|d_{3\alpha\beta}D_{+\alpha}D_{8\beta}|n\rangle$ are to be 
computed in the RRA approach with full inclusion of flavor symmetry breaking
effects.

\begin{figure}[b]
\centerline{
\includegraphics[width=3.5cm,height=5.5cm,angle=270]{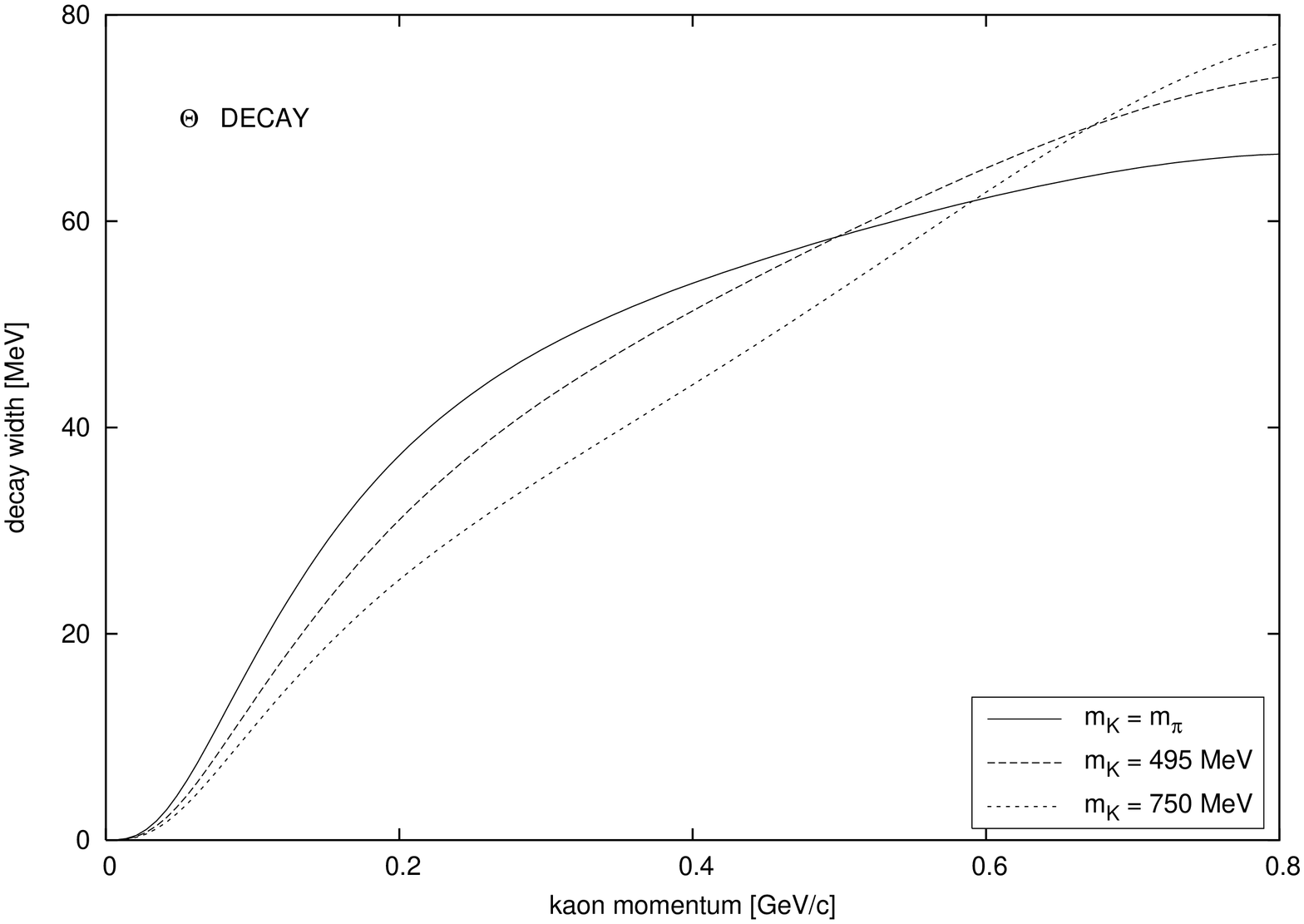}\hspace{2cm}
\includegraphics[width=3.5cm,height=5.5cm,angle=270]{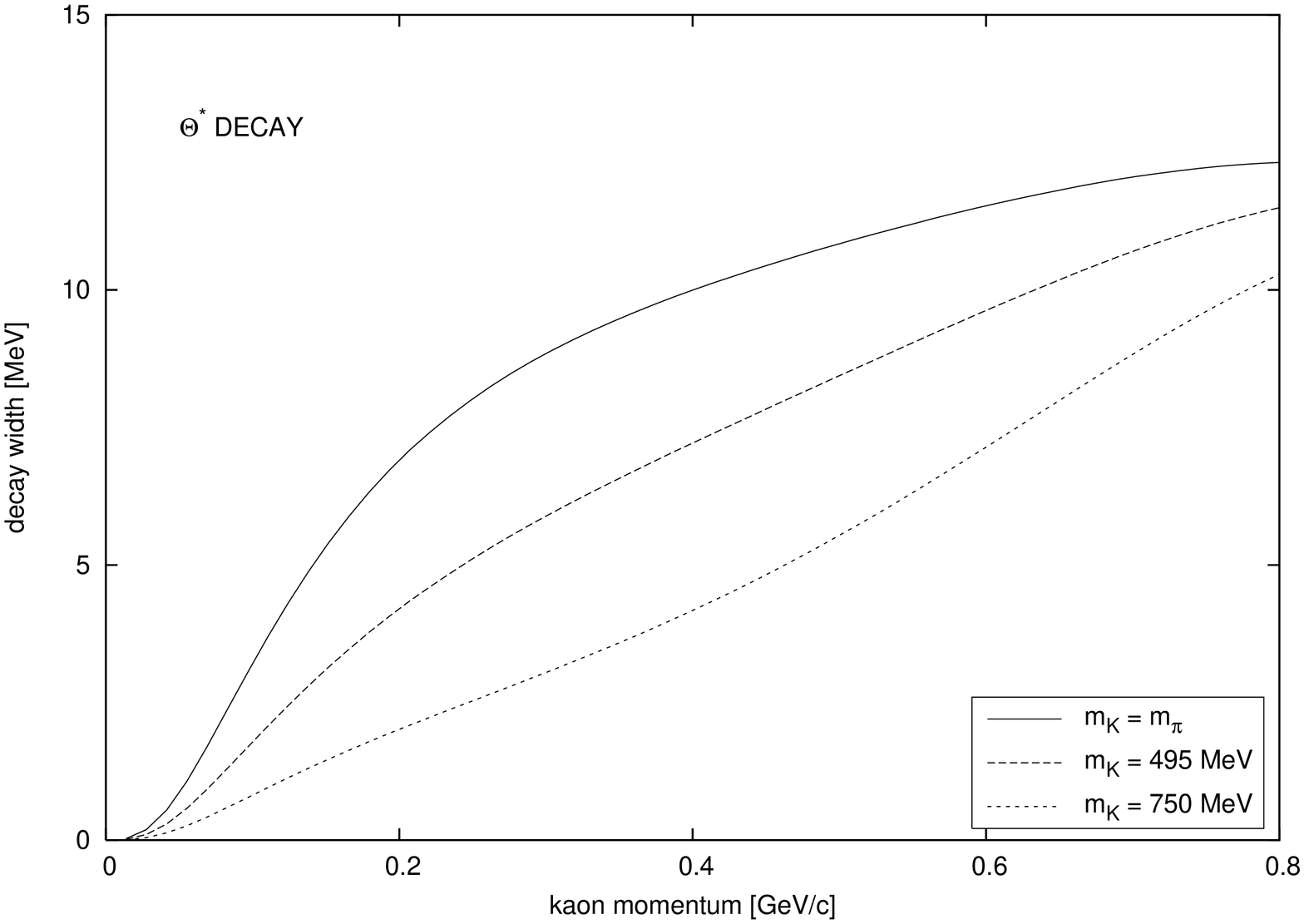}}
\caption{\label{fig3}Model prediction for the
width, $\Gamma(\omega)$ of $\Theta^+$ (left) and $\Theta^{*+}$ (right)
for $N_C=3$ as function of the momentum
$k=\sqrt{\omega^2-m_K^2}$ for three values of the kaon mass.
Note the unequal scales.}
\end{figure}

This width function is shown (for $N_C=3$) in figure~\ref{fig3} for 
$\Theta$ and its isovector partner $\Theta^*$. The latter merely 
requires the appropriate modification of the matrix elements in 
eq.~(\ref{ddr}). Most importantly, the $k^3$ behavior of the width
function, as suggested by the model, eq.~(\ref{eq:yukawa}) is 
reproduced only right above threshold, afterwards it levels off.
Second, and somewhat surprising, the width of the non--ground 
state pentaquark is smaller than that of the lowest lying 
pentaquark. Our particular model yields
$\Gamma_\Theta\approx{40}{\rm MeV}$ and 
$\Gamma_{\Theta^*}\approx{20}{\rm MeV}$. We note that there
are certainly model ambiguities in these results.

\section{Conclusions}

We have discussed the chiral soliton model approach to $KN$ scattering 
in the $S=+1$ channel which contains the potential~\cite{Hi07} $\Theta^+$ 
pentaquark, a state predicted as a flavor rotational excitation of the 
soliton. Though the exactly known large~$N_C$ phase shift suggests otherwise, 
the $\Theta$ emerges as a genuine resonance. Our central result is the width 
function for $\Theta\to KN$.  In the flavor symmetric case it contains only 
a \emph{single} collective coordinate operator and is thus very different 
from estimates that extract an effective Yukawa coupling from the axial 
current matrix element~\cite{Di97}. Since our approach matches the 
exact large $N_C$ result, we must conclude that those axial current scenarios 
are erroneous~\cite{We07} and that the cancellation among contributions to 
this matrix element is an invalid argument for a small pentaquark 
width.

\section*{Acknowledgments}

This key note is based on a collaboration with H. Walliser,
whose contribution is highly appreciated.
I am very grateful to the organizers of this workshop for the 
invitation to contribute to the proceedings despite 
I was unable to participate.

\end{document}